\documentclass[12pt]{article}
\usepackage[dvips]{color}
\usepackage{epsfig}
\usepackage{amsmath}
\usepackage{graphicx}
\def\Box{\hbox{$\rlap{$\sqcup$}\sqcap$}}
\begin{document}
\setcounter{page}{1}

\pagestyle{plain} \vspace{1cm}

\begin{center}
\Large{\bf Tachyonic Teleparallel Dark Energy}\\
\small \vspace{1cm} {\bf A. Banijamali
\footnote{a.banijamali@nit.ac.ir}} and {\bf B. Fazlpour
\footnote{b.fazlpour@umz.ac.ir}}\\
\vspace{0.5cm}   {\it Department of Basic Sciences, Babol University
of Technology, Babol, Iran\\}
\end{center}
\vspace{1.5cm}
\begin{abstract}
Teleparallel gravity is an equivalent formulation of general
relativity in which instead of the Ricci scalar $R$, one uses the
torsion scalar $T$ for the Lagrangian density. Recently teleparallel
dark energy has been proposed by Geng et al. in (Geng et al., 2011).
They have added quintessence scalar field, allowing also a
non-minimal coupling with gravity in the Lagrangian of teleparallel
gravity and found that such a non-minimally coupled quintessence
theory has a richer structure than the same one in the frame work of
general relativity. In the present work we are interested in
tachyonic teleparallel dark energy in which scalar field is
responsible for dark energy in the frame work of torsion gravity. We
find that such a non-minimally coupled tachyon gravity can realize
the crossing of the phantom divide line for the effective equation
of state. Using the numerical
calculations we display such a behavior of the model explicitly.\\


\end{abstract}
\newpage
\section{Introduction}
Recent cosmological observations from supernovae Ia (Perlmutter et
al. 1999; Riess et al., 1998), cosmic microwave background radiation
(Spergel et al., 2003; Komatsu et al., 2009; Komatsu et al., 2010),
large scale structure (Tegmark et al., 2004; Seljak et al., 2005),
baryon acoustic oscillations (Eisenstein et al., 2005) and weak
lensing (Jain $\&$ Taylor, 2003) have revealed that our universe is
in accelerated expansion phase and it began this acceleration at the
near past. In order to explain the late time cosmic acceleration one
can use dark energy or dark gravity approaches. In the dark gravity
approach, one modifies the left-hand side of the Einstein equation
to obtain a modified gravity theory. The simplest model of this
category is the well-known $f(R)$ gravity in which the Ricci scalar
in the Einstein-Hilbert action replaces by a general function of the
Ricci scalar (Nojiri $\&$ Odintsov, 2006; De Felice $\&$ Tsujikawa,
2010; Sotiriou $\&$ Faraoni, 2010). In the second approach we
introduce an exotic energy component with negative pressure called
dark energy in the right-hand side of the Einstein equation in the
frame work of
general relativity (for a review see (Copeland et al., 2006)).\\
The simplest candidate of dark energy is a tiny positive
time-independent cosmological constant $\Lambda$ with the equation
of state $\omega=-1$ (Weinberg, 1989; Sahi $\&$ Starobinsky, 2000;
Peebles $\&$ Ratra, 2003; Padmanabhan, 2003). However, it suffers
from two serious theoretical problems, i.e., the cosmological
constant problem (why $\Lambda$ is about 120 orders of magnitude
smaller than its natural expectation value?) and the coincidence
problem (why are we living in an epoch in which the dark energy
density and the dust matter energy are comparable?). As a solution
of these problems various dynamical dark energy models have been
proposed. The dynamical nature of dark energy can originate from
various scalar fields such as quintessence (Wetterich, 1988; Ratra
$\&$ Peebles, 1988; Caldwell et al., 1998), phantom (a scalar field
with negative kinetic energy) and also tachyon scalar field
(Alexander, 2002; Mazumdar et al., 2001; Gibbons, 2002; Garousi et
al., 2005; Copeland et al., 2005; Sen, 1999; Bergshoeff et al.,
2000; Kluson, 2000). Holographic dark energy is another proposal for
dynamical dark energy models (Setare, 2007a, b, c).\\
For a single minimally coupled scalar field model, it has been shown
that the effective equation of state cannot cross the phantom divide
line ($\omega=-1$) confirmed by cosmic observations (Zhao et al.,
2005; Caldwell $\&$ Doran, 2005; Hu, 2005; Feng et al., 2005).
Indeed, it is shown in (Vikman, 2005) that the transition from
$\omega>-1$ to $\omega<-1$ (or vice versa) of dark energy described
by general scalar field Lagrangian $\mathcal{L}(\phi,
\partial_{\mu}\phi)$ is impossible. Thus in order to explain the
transition under the minimal assumptions of the non-kinetic
interaction of dark energy and other matter one should suppose that
the dark energy was subdominating and described by a nonlinear in
$(\partial_{\mu}\phi)^{2}$ Lagrangian. So, some nonlinear (or
probably quantum) physics must be invoked to explain the value
$\omega<-1$ in models with one scalar field. Also, the models with a
combination of phantom and quintessence called quintom have been
proposed (Feng et al., 2005; Elizalde et al., 2004; Guo et al.,
2005; Wu $\&$ Yu, 2005; Cai et al., 2007a; Chimento et al., 2009;
Cai et al., 2010; Wei et al., 2005a, b; Alimohammadi $\&$ Mohseni
Sadjadi, 2006; Zhao $\&$ Zhang, 2006; Wei et al., 2007). Scalar
field model with non-linear kinetic terms (Nojiri et al., 2005; Sami
et al., 2005; Leith $\&$ Neupane, 2007; Nojiri et al., 2006; Kovitso
$\&$ Mota, 2007; Setare $\&$ Saridakis, 2008; Sanyal, 2007) or a
non-linear higher derivative one (Vikman, 2005), brane world models
(Deffayet et al., 2002; Setare, 2006), string-inspired models
(Cataldo $\&$ Chimento, 2007), modified gravity models and
non-minimally coupled scalar field models where scalar field couples
with the Ricci scalar, Gauss-Bonnet invariant or modified $f(R)$
gravity have also been constructed to realize the crossing of the
phantom divide line (Caldwell, 2002; Nojiri $\&$ Odintsov, 2003;
Perivolaropoulos, 2005; Sadeghi et al., 2009)
(for a detailed review, see (Padmanabhan, 2003)).\\
Furthermore, the teleparallel equivalent of the Einstein general
relativity (Einstein, 1928; Hayashi $\&$ Shirafuji, 1979) is an
equivalent formulation of classical gravity where one uses the
Weitzenbock connection, which has no curvature but torsion, rather
than the torsionless Levi-Civita connection. The dynamical objects
in this formulation are four linearly independent vierbeins. The
advantage of this framework is that the torsion is formed solely
from products of first derivative of the tetrad.\\
Recently, following the $f(R)$ modified gravity, generalizations of
teleparallel gravity have been proposed in Refs. (Bengochea $\&$
Ferraro, 2009; Wu $\&$ Yu, 2010a, b; Linder, 2010;Wu $\&$ Yu, 2011;
Chen et al., 2011; Bengochea, 2011; Yang, 2011; Zheng $\&$ Huang,
2011; Li et al., 2011). That is $f(T)$ gravity, where $f$ is a
general function and $T$ is the Lagrangian of teleparallel gravity.
The interesting feature of $f(T)$ theories is that their equations
are second order in derivatives and therefore, they can use to
account for the accelerated expansion of the universe and remain
free of pathologies.\\
A number of attempts in studying $f(T)$ gravity are as follows:
Finite-time future singularities in $f(T)$ gravity models have been
considered in (Setare $\&$ Houndjo, 2012). The cosmological
perturbations in such a theory have been investigated in (Dent et
al., 2010; Zheng $\&$ Huang, 2010) and local Lorentz invariance in
this
context has been examined in (Sotiriou et al., 2010; Li et al., 2010).\\
Very recently the teleparallel dark energy in which a usual scalar
field in the action of teleparallel gravity is responsible for dark
energy, has been proposed in (Geng et al., 2011). Geng et al. have
considered quintessence as dark energy component and showed that
such a scenario is completely equivalent to the standard
quintessence if scalar field minimally coupled to gravity. Motivated
by the similar one in the framework of general relativity, they have
also investigated the case that the scalar field non-minimally
coupled to the torsion scalar in the framework of teleparallel
gravity. They have found that such a theory has a richer structure
than the same one in the framework of general relativity. The richer
structure of non-minimally coupled quintessence with torsion gravity
is due to exhibiting quintessence-like or phantom-like behavior, or
experiencing the phantom divide crossing in this theory.\\
In this paper we are interested  in teleparallel dark energy where
instead of quintessence one uses the tachyon scalar field. The
tachyon field in the world volume theory of the open string
stretched between a D-brane and an anti-D-brane or a non-BPS D-brane
plays the role of scalar field in the context of string theory
(Alexander, 2002; Mazumdar et al., 2001; Gibbons, 2002). What
distinguishes the tachyon Lagrangian from the standard Klein-Gordan
form for scalar field is that the tachyon action has a non-standard
type namely, Dirac-Born-Infeld form (Garousi et
al., 2005; Copeland et al., 2005; Sen, 1999; Bergshoeff et al., 2000; Kluson, 2000).\\
An outline of the present work is as follows: In section 2 we
introduce tachyonic teleparallel dark energy model in which tachyon
field plays the role of scalar field and the non-minimal coupling
between scalar field and torsion scalar is also present in the
action. Then, we derive the field equations as well as the energy
density and pressure in order to study the cosmological behavior of
the model. We obtain the conditions required for phantom divide
crossing of our model in section 3 and then we show that such a
crossing can be satisfied numerically. Section 4 is devoted to our
conclusions.\\

\section{The Model}
In this section let us first briefly review the key ingredients of
teleparallel gravity. To this end we follow the Refs. (Einstein,
1928; Hayashi $\&$ Shirafuji, 1979; Bengochea $\&$ Ferraro, 2009; Wu
$\&$ Yu, 2010a, b; Linder, 2010; Wu $\&$ Yu, 2011). In the
teleparallelism, orthonormal tetrad components $e_{i}(x^{\mu})$ are
used, where an index $i$ runs over $0,1,2,3$ for the tangent space
at each point $x^{\mu}$ of the manifold. Their relation to the
metric $g_{\mu\nu}$ is given by
\begin{equation}
g_{\mu\nu}=\eta_{ij}e^{i}_{\mu}e^{j}_{\nu},
\end{equation}
where $\mu$ and $\nu$ are coordinate indices on the manifold, also
running over $0,1,2,3$. Instead of the Ricci scalar $R$ for the
Lagrangian density in general relativity, the teleparallel
Lagrangian density is described by the torsion scalar $T$, defined
as
\begin{equation}
T\equiv S_{\rho}^{\,\,\,\,\mu\nu}T^{\rho}_{\,\,\,\,\mu\nu}
\end{equation}
where the torsion $T^{\rho}_{\,\,\,\,\mu\nu}$, contorsion
$K^{\mu\nu}_{\,\,\,\,\,\rho}$ and $S_{\rho}^{\,\,\,\,\mu\nu}$,
defined as
\begin{equation}
T^{\rho}_{\,\,\,\,\mu\nu}\equiv
e_{i}^{\,\rho}\big(\partial_{\mu}e^{i}_{\nu}-\partial_{\nu}e^{i}_{\mu}\big),
\end{equation}
\begin{equation}
K^{\mu\nu}_{\,\,\,\,\,\rho}\equiv
-\frac{1}{2}\big(T^{\mu\nu}_{\,\,\,\,\,\rho}-T^{\nu\mu}_{\,\,\,\,\,\rho}-T^{\,\,\,\,\mu\nu}_{\rho}\big),
\end{equation}
\begin{equation}
S_{\rho}^{\,\,\,\,\mu\nu}\equiv
\frac{1}{2}\big(K^{\mu\nu}_{\,\,\,\,\,\rho}+\delta_{\,\rho}^{\,\mu}T^{\theta\nu}_{\,\,\,\,\,\theta}-
\delta_{\,\rho}^{\,\nu}T^{\theta\mu}_{\,\,\,\,\,\theta}\big).
\end{equation}
In summary, the relevant action of teleparallel gravity is
\begin{equation}
S=\int d^{4}xe \Big[\frac{T}{2\kappa^{2}}+\mathcal{L}_{m}\Big],
\end{equation}
where $e=det(e_{\,\mu}^{\,i})=\sqrt{-g}$ and $\kappa^{2} = 8\pi G =
\frac{1}{M_{Pl}^{2}}$ while $G$ is a bare gravitational constant and
$M_{Pl}$ is a reduced Planck mass. Since the above action cannot
lead to an accelerated universe, one has to generalize the action
(6) by using the following two approaches: the first is to replace
$T$ by a general function $f(T)$ (Geng et al., 2011; Unzicker $\&$
Case, 2005; Chattopadhyay $\&$ Debnath, 2011; Sharif $\&$ Rani,
2011; Wei et al., 2011) in analogy to $f(R)$ extension of general
relativity. The second way is to add a scalar field responsible for
dark energy in action (6), allowing a non-minimal coupling between
it and gravity. We focus on the second
approach and utilize the tachyon scalar field as dark energy.\\
Our starting action is as follows,
\begin{equation}
S=\int d^{4}xe
\Big[\frac{T}{2\kappa^{2}}+f(\phi)T-V(\phi)\sqrt{1+g^{\mu\nu}\partial_{\mu}\phi\partial^{\nu}\phi}
+\mathcal{L}_{m}\Big],
\end{equation}
where $f(\phi)$ is an arbitrary function of tachyon scalar field and
it is responsible for non-minimal coupling between tachyon and
gravity. The last term in the above action is the Born-Infeld type
action for tachyon field, where $V(\phi)$ is the tachyonic potential
which is bounded and reaching its minimum asymptotically. Similar to
the standard non-minimal tachyon in general relativity where the
scalar field couples to the Ricci scalar, in here the non-minimal
coupling will be between the torsion and the scalar field.\\
In Refs. (Cai et al., 2007b; Sadeghi et al., 2008) the Authors have
been added an extra higher derivative term $T\Box T$ in the square
root part of the tachyonic Lagrangian and showed that crossing of
the phantom divide occurs before
reaching the tachyon potential asymptotically to its minimum.\\
Considering a spatially-flat Friedmann-Robertson-Walker (FRW)
metric,
\begin{eqnarray}
ds^{2}=-dt^{2}+a^{2}(t)(dr^{2}+r^{2}d\Omega^{2}),
\end{eqnarray}
and a vierbein choice of the form $e^{i}_{\mu}=diag(1,a,a,a)$ and a
homogeneous scalar field $\phi$, the corresponding Friedmann
equations are given by,
\begin{equation}
H^{2}=\frac{\kappa^{2}}{3}\big(\rho_{\phi}+\rho_{m}\big),
\end{equation}
\begin{equation}
\dot{H}=-\frac{\kappa^{2}}{2}\big(\rho_{\phi}+P_{\phi}+\rho_{m}+P_{m}\big),
\end{equation}
where $H=\frac{\dot{a}}{a}$ is the Hubble parameter, $a$ is the
scale factor, a dot stands for the derivative with respect to cosmic
time $t$, $\rho_{m}$ and $P_{m}$ are the matter energy density and
pressure respectively, satisfying the equation
$\dot{\rho}_{m}+3H(1+\omega_{m})\rho_{m}=0$, with
$\omega_{m}=\frac{P_{m}}{\rho_{m}}$ the matter equation of state
parameter. The effective energy density and pressure of tachyonic
teleparallel dark energy are given by,
\begin{equation}
\rho_{\phi}=\frac{V(\phi)}{\sqrt{1-\dot{\phi}^{2}}}-6 H^{2}f(\phi),
\end{equation}
and
\begin{equation}
P_{\phi}=-V(\phi)\sqrt{1-\dot{\phi}^{2}}+2\big(3H^{2}+2\dot{H}\big)f(\phi)+2
H f'(\phi)\dot{\phi}\big(1+\frac{1}{\sqrt{1-\dot{\phi}^{2}}}\big),
\end{equation}
where a prime denotes derivative with respect to $\phi$. Note that
for a FRW universe it is easy to find $T=-6H^{2}$.\\
The equation of motion of the scalar field in FRW background (8)
reads
\begin{equation}
\frac{\ddot{\phi}}{1-\dot{\phi}^{2}}+3H\dot{\phi}+\frac{V'(\phi)}{V(\phi)}+\frac{6H^{2}f'(\phi)}{V(\phi)}=0.
\end{equation}
In fact the above relation expresses the energy conservation
equation
\begin{equation}
\dot{\rho}_{\phi}+3H(1+\omega_{\phi})\rho_{\phi}=0.
\end{equation}
From equation (13) one can obtain the following useful relation for
the Hubble parameter $H$,
\begin{equation}
H=\frac{1}{4}\Bigg[-\frac{\dot{\phi}V(\phi)}{f'(\phi)}\pm\frac{1}{f'(\phi)}\sqrt{\dot{\phi}^{2}V^{2}(\phi)-\frac{8}{3}
f'(\phi)V'(\phi)+\frac{8}{3}\frac{f'(\phi)V(\phi)\ddot{\phi}}{(\dot{\phi}^{2}-1)}}\,\Bigg].
\end{equation}
Next, we are going to explore the effects of non-minimal
teleparallel dark energy on the cosmological evolution of equation
of state (EoS) and see how the present model can be used to realize
a crossing of phantom divide $\omega=-1$ for two possible roots of $H$.\\
\section{The $\omega=-1$ Crossing}
Let us proceed in studying cosmological implications of the present
model. For doing so, we begin with the EoS of teleparallel dark
energy $\omega_{\phi}=\frac{P_{\phi}}{\rho_{\phi}}$ and assume that
the matter content of the univers to be dust matter
$(\omega_{m}\approx 0)$. Now one derives the following relation
using (11) and (12),
\begin{equation}
\rho_{\phi}+P_{\phi}=\frac{V(\phi)
\dot{\phi}^{2}}{\sqrt{1-\dot{\phi}^{2}}}+2
Hf'(\phi)\dot{\phi}\big(1+\frac{1}{\sqrt{1-\dot{\phi}^{2}}}\big)+4\dot{H}f(\phi).
\end{equation}
When $\omega_{\phi}$ goes to $-1$ the above expression should be
zero because we have
$\rho_{\phi}+P_{\phi}=3H(1+\omega_{\phi})\rho_{\phi}$. So, to
fulfilled such a requirement, one leads to the condition,
\begin{equation}
\dot{\phi}\Big(\frac{V(\phi)\dot{\phi}}{\sqrt{1-\dot{\phi}^{2}}}+2
Hf'(\phi)\big(1+\frac{1}{\sqrt{1-\dot{\phi}^{2}}}\big)\Big)=-4\dot{H}f(\phi).
\end{equation}
Moreover, in order to $\omega_{\phi}$ crosses the phantom divide
line, we have to check $\frac{d}{dt}(\rho_{\phi}+P_{\phi})\neq 0$ at
the crossing point,
$$\frac{d}{dt}(\rho_{\phi}+P_{\phi})=\frac{V(\phi)
\dot{\phi}^{3}}{\sqrt{1-\dot{\phi}^{2}}}+\frac{2V(\phi)
\dot{\phi}\ddot{\phi}}{\sqrt{1-\dot{\phi}^{2}}}+\frac{V(\phi)
\dot{\phi}^{3}\ddot{\phi}}{\big(1-\dot{\phi}^{2}\big)^{\frac{3}{2}}}+4\ddot{H}f(\phi)
+2\dot{H}f'(\phi)\dot{\phi}\big(3+\frac{1}{\sqrt{1-\dot{\phi}^{2}}}\big)$$
\begin{equation}
+2H\Bigg[f'(\phi)\ddot{\phi}\big(1+\frac{1}{(1-\dot{\phi}^{2})^{\frac{3}{2}}}\big)+f''(\phi)\dot{\phi}^{2}
\big(1+\frac{1}{\sqrt{1-\dot{\phi}^{2}}}\big)\Bigg].
\end{equation}
Next, putting the condition (17) into (18), yields to
$$\frac{d}{dt}(\rho_{\phi}+P_{\phi})=\frac{V(\phi)
\dot{\phi}^{3}}{\sqrt{1-\dot{\phi}^{2}}}\Big(1+\frac{\ddot{\phi}}{1-\dot{\phi}^{2}}\Big)
+4\ddot{H}f(\phi)
+2\dot{H}\Bigg[f'(\phi)\dot{\phi}\big(3+\frac{1}{\sqrt{1-\dot{\phi}^{2}}}\big)-4f(\phi)\frac{\ddot{\phi}}{\dot{\phi}}\Bigg]$$
\begin{equation}
+2H\Bigg[f''(\phi)\dot{\phi}^{2}\big(1+\frac{1}{\sqrt{1-\dot{\phi}^{2}}}\big)-
f'(\phi)\ddot{\phi}\Big(1+\frac{(1-2\dot{\phi}^{2})}{(1-\dot{\phi}^{2})^{\frac{3}{2}}}\Big)\Bigg].
\end{equation}
Now, the condition (17) tell us that a necessary condition to obtain
phantom divide crossing is $\dot{\phi}\neq 0$. We remark that the
condition (17) is a necessary condition and not enough condition for
crossing over $-1$, but one immediately concludes from (19), if
$\dot{\phi}\neq 0$ then (19) will be non-zero and the enough
condition fulfilled. In order to show that non-minimal tachyonic
teleparallel dark energy can indeed realized the phantom divide
crossing more transparently we solve the cosmological system
numerically using specific examples for coupling function $f(\phi)$
and tachyonic potential $V(\phi)$. \\
In figure 1, we have plotted the evolution of equation of state
parameter for two kinds of tachyonic potentials, i.e. exponential
potential $V(\phi)=V_{0}e^{-\alpha \phi^{2}}$ and inverse square
potential $V(\phi)=\frac{V_{0}}{\phi^{2}}$, and for positive sign in
expression (15). We mention that in the numerical calculations we
have considered the non-minimal coupling function to be
$f(\phi)=\phi^{2}$. Such a non-minimal coupling function is the same
as that considered in (Geng et al., 2011). These plots clearly show
that crossing of the phantom divide line can be realized from a
non-phantom phase $(\omega > -1)$ to a phantom phase $(\omega < -1)$
in non-minimal tachyonic teleparallel dark energy. It is interesting
to note that a crossing from $(\omega > -1)$ to $(\omega < -1)$ is
consistent with the recent cosmological observational data (Alam et
al., 2004; Nesseris $\&$ Perivolaropoulos, 2007; Wu $\&$ Yu, 2006;
Alam et al., 2007; Jassal et al., 2010) and also such a crossing is
in agreement with that in the context of $f(T)$ gravity models
(Bamba et al., 2011) but in contrast with the viable $f(R)$ gravity
models where the phantom divide line is crossed the other way around
(Bamba et al., 2010a, b). In figure 2, we have depicted the
$\omega$-evolution for the same choices as in figure 1 but for minus
sign in (15). One can see that in this case the crossing of $-1$
cannot be happen for $V(\phi)=\frac{V_{0}}{\phi^{2}}$ and for
exponential potential such a crossing is from phantom phase to the
non-phantom one. So, by considering the above discussion it seems
that the exponential potential is a better choice in studying the
phantom divide crossing cosmology in the context of non-minimal
teleparallel tachyonic dark energy.\\

\begin{figure}[htp]
\begin{center}
\includegraphics{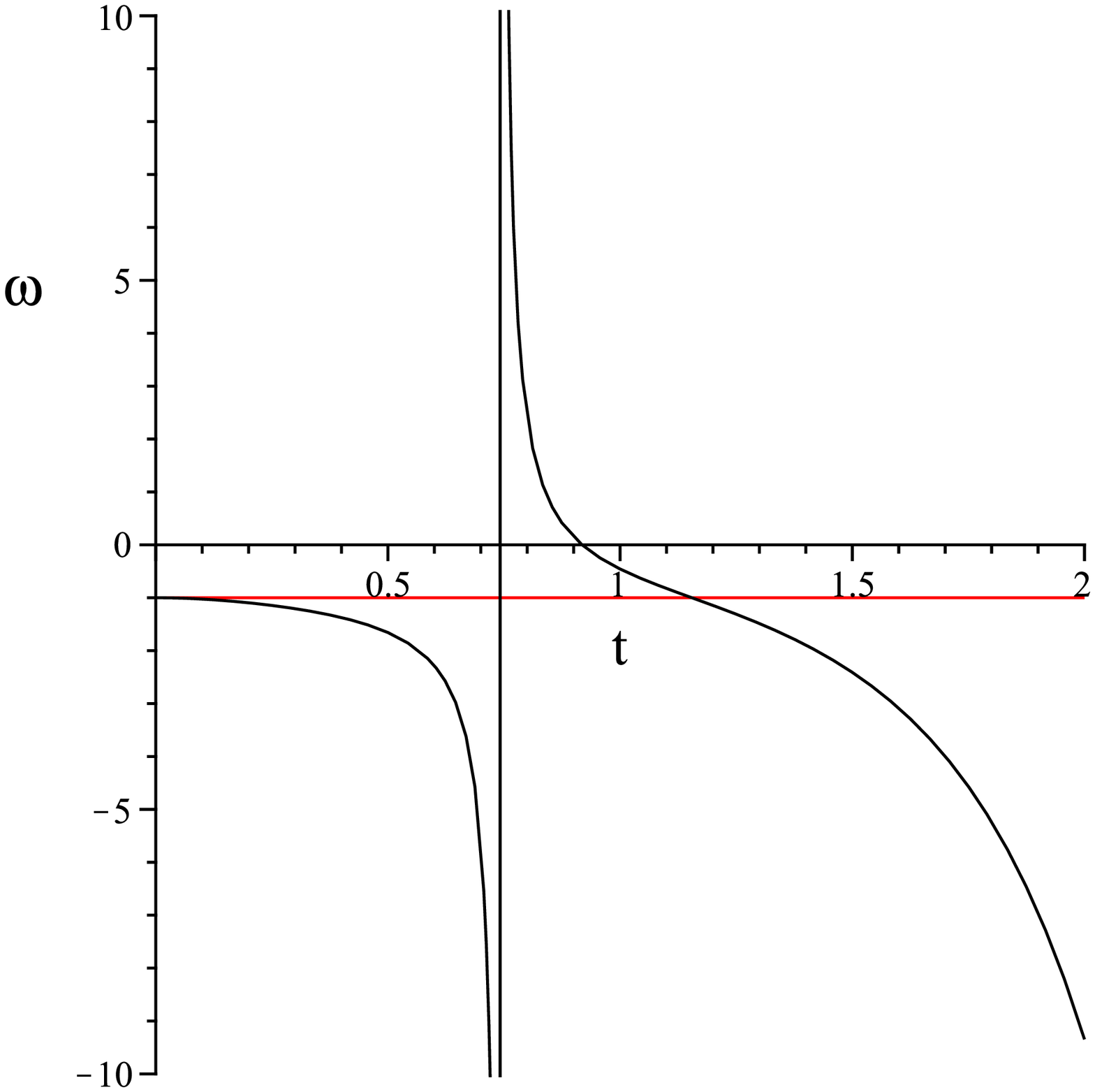} \vspace{8.5cm}\includegraphics{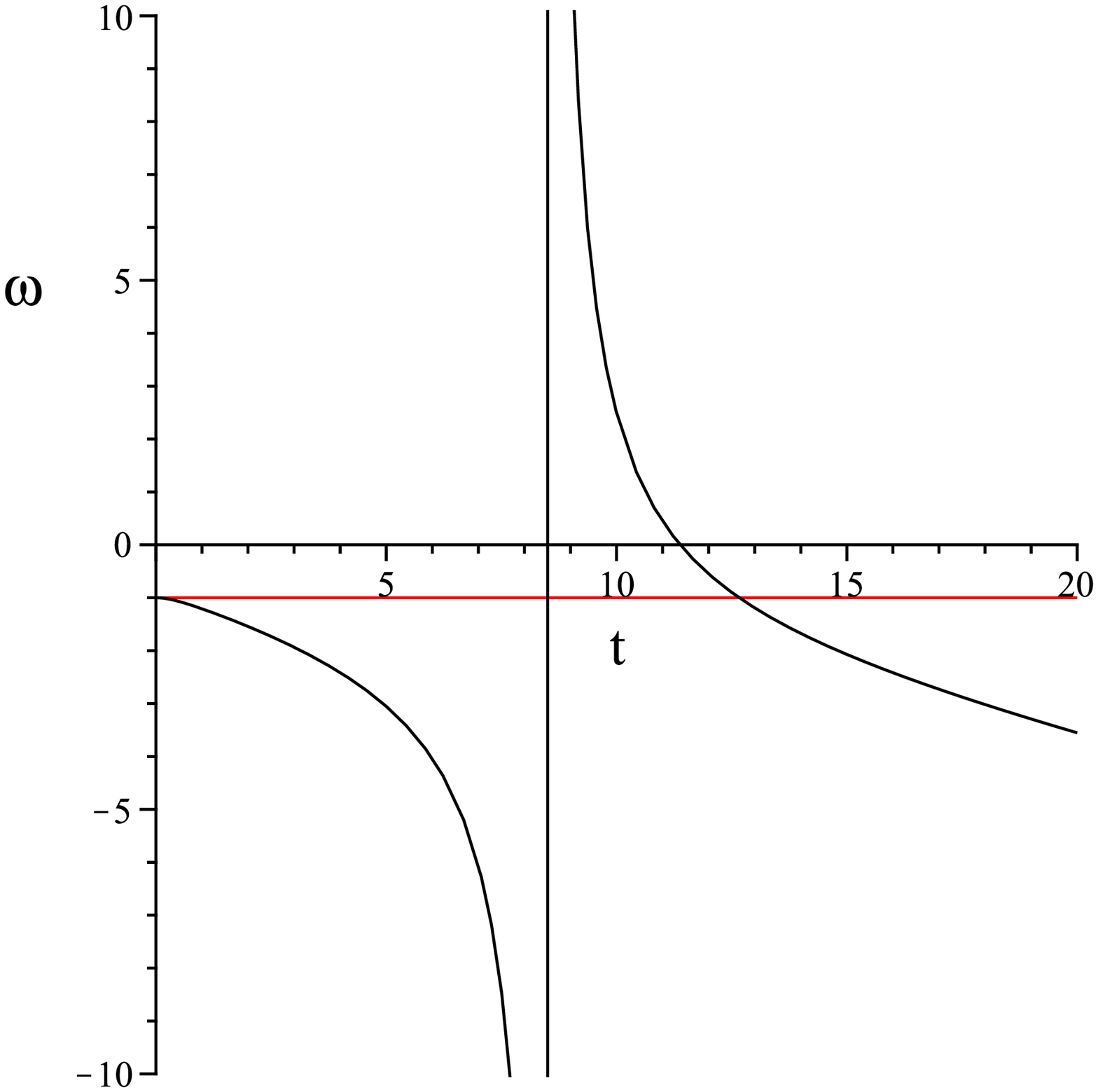}
\end{center}
 \caption{\small { Plots of the evolution of the
EoS versus $t$ for positive root of $H$ (left for the potential
$V(\phi)=V_{0}e^{-\alpha \phi^{2}}$ and right for the potential
$V(\phi)=\frac{V_{0}}{\phi^{2}}$) , $f(\phi)=b\phi^{n}$, (with
$b=1$, $n=2$, $V_{0}=4$, $\phi_{0}=0.5$ and $\alpha=5$).}}
\end{figure}

\begin{figure}[htp]
\begin{center}
\includegraphics{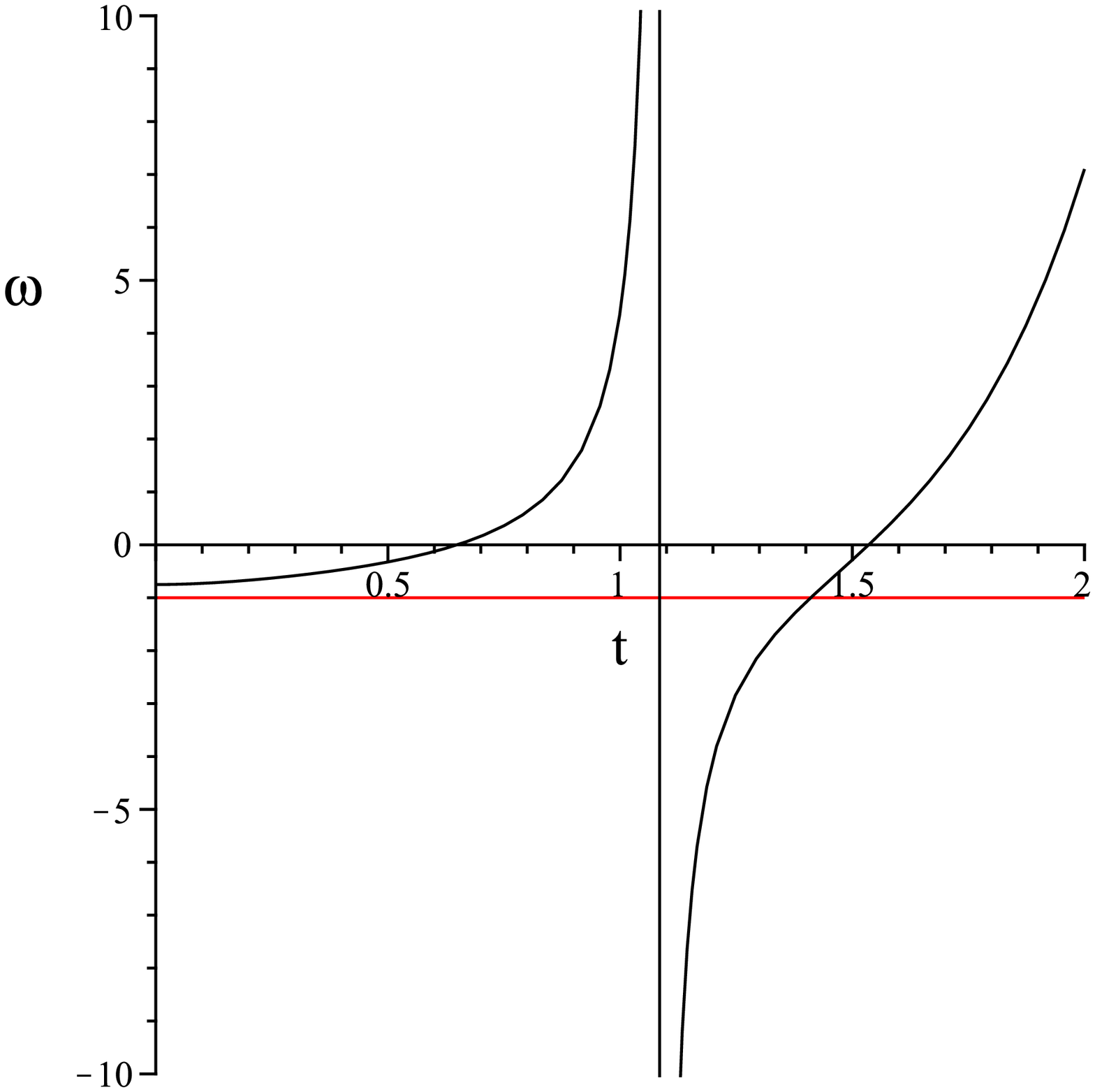} \vspace{7cm}\includegraphics{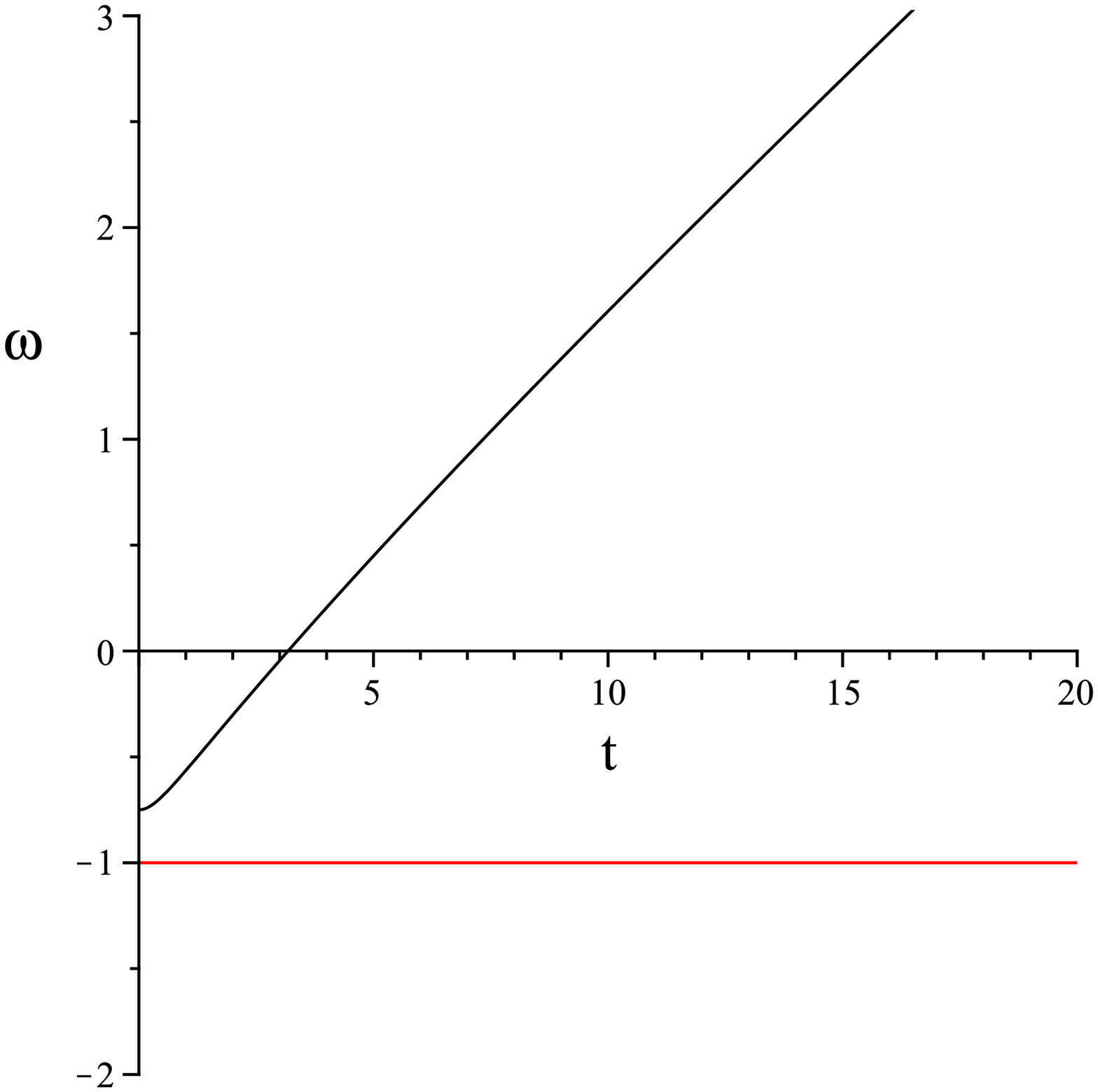}
\end{center}
 \caption{\small { Plots of the evolution of the
EoS versus $t$ for negative root of $H$ (left for the potential
$V(\phi)=V_{0}e^{-\alpha \phi^{2}}$ and right for the potential
$V(\phi)=\frac{V_{0}}{\phi^{2}}$) , $f(\phi)=b\phi^{n}$, (with
$b=1$, $n=2$, $V_{0}=4$, $\phi_{0}=0.5$ and $\alpha=5$).}}
\end{figure}
\newpage
\section{Conclusion}
The teleparallel dark energy scenario (Geng et al., 2011) is based
on the teleparallel equivalent of general relativity. In such a
model one utilizes a scalar field, in which the dark energy sector
is attributed, allowing also for a non-minimal coupling between the
field and the torsion scalar. We have proposed thachyonic
teleparallel dark energy model where tachyon field played the role
of scalar field. A remarkable feature of the our model is that it
realizes the phantom divide line crossing, a cosmologically observed
phenomena (Alam et al., 2004; Nesseris $\&$ Perivolaropoulos, 2007;
Wu $\&$ Yu, 2006; Alam et al., 2007; Jassal et al., 2010). By
studying the evolutionary curves of $\omega_{\phi}$ for exponential
and inverse square tachyonic potential, we found that
$\omega_{\phi}$ can cross the $-1$ line as it was shown in figures 1
and 2. Furthermore, we also found that at the crossing point the
time derivative of tachyon field, $\dot{\phi}$ should be non-zero in
order to $\omega_{\phi}$ crosses over $-1$. In addition, considering
the positive sign in equation (15) leads to a crossing from
non-phantom phase to the phantom phase for both exponential and
inverse square tachyonic potentials and such a behavior is
consistent with the observational data (Alam et al., 2004; Nesseris
$\&$ Perivolaropoulos, 2007; Wu $\&$ Yu, 2006; Alam et al., 2007;
Jassal et al., 2010). In the other side taking the negative sign in
(15) and exponential form for tachyon potential result in phantom
divide crossing similar to that in viable $f(R)$ models i.e from
$(\omega < -1)$ to $(\omega
> -1)$, while for inverse square potential the $(\omega = -1)$
crossing does not
occur.\\
In summary, the model (7) has the property of crossing the
cosmological constant boundary, only with single tachyon field which
is coupled non-minimally with torsion gravity. In contrast to Refs.
(Cai et al., 2007b; Sadeghi et al., 2008; Li et al., 2005) here we
do not need to add an extra term or a dimension-6 operator $\Box T
\Box T$ in the Lagrangian of tachyon field.\\

\end{document}